\documentclass[a4paper,english,conference]{IEEEtran}
\IEEEoverridecommandlockouts
\usepackage{latexsym}
\usepackage{verbatim}
\usepackage{amsfonts}
\usepackage{amsbsy}
\usepackage{amssymb}
\usepackage{flushend,cuted}
\usepackage{amsmath,amssymb}
\usepackage{times}
\usepackage{graphicx}
\usepackage{enumerate}
\usepackage[usenames]{color}
\usepackage[dvips]{pstcol}
\usepackage{epstopdf}
\usepackage{cite}
\usepackage{amsmath}
\usepackage{amssymb}
\usepackage{amsfonts}
\usepackage{graphicx}
\usepackage{epsfig}
\usepackage{psfrag}
\usepackage{xcolor}
\usepackage{amsfonts, bm}
\usepackage{epstopdf}
\usepackage{cite}
\usepackage{color}
\usepackage{xcolor}
\usepackage{subfig}
\usepackage{algorithm}
\usepackage{algorithmicx}
\usepackage{algpseudocode}
\usepackage{amsmath}
\usepackage{multirow}
\input epsf
\begin{document}

\title{One-shot Learning for Channel Estimation in Massive MIMO Systems}

\author{\IEEEauthorblockN{ Kai Kang \IEEEauthorrefmark{1}, Qiyu Hu \IEEEauthorrefmark{1}, Yunlong Cai \IEEEauthorrefmark{1}, and Yonina C. Eldar\IEEEauthorrefmark{2} }
\IEEEauthorblockA{\IEEEauthorrefmark{1} College of Information Science and Electronic Engineering, Zhejiang University, Hangzhou, China }
\IEEEauthorblockA{\IEEEauthorrefmark{2} Department of Mathematics and Computer Science, Weizmann Institute of Science, Rehovot 7610001, Israel  \\ E-mail: \{kangkai, qiyhu, ylcai,\}@zju.edu.cn, yonina.eldar@weizmann.ac.il}
\thanks{Zhejiang Provincial Key Laboratory of Information Processing, Communication and Networking (IPCAN), Hangzhou 310027, China. This work was supported in part by the National Natural Science Foundation of China under Grants 61971376, U22A2004, and 61831004 and supported by the Fundamental Research Funds for the Central Universities 226-2022-00195.}
}

\maketitle

\begin{abstract}
	In conventional supervised deep learning based channel estimation algorithms, a large number of training samples are required for offline training. However, in practical communication systems, it is difficult to obtain channel samples for every signal-to-noise ratio (SNR). Furthermore, the generalization ability of these deep neural networks (DNN) is typically poor. In this work, we propose a one-shot self-supervised learning framework for channel estimation in multi-input multi-output (MIMO) systems. The required number of samples for offline training is small and our approach can be directly deployed to adapt to variable channels. Our framework consists of a traditional channel estimation module and a denoising module. The denoising module is designed based on the one-shot learning method Self2Self and employs Bernoulli sampling to generate training labels. Besides,we further utilize a blind spot strategy and dropout technique to avoid overfitting. Simulation results show that the performance of the proposed one-shot self-supervised learning method is very close to the supervised learning approach while obtaining improved generalization ability for different channel environments. 
\end{abstract}
\begin{IEEEkeywords}
	Channel estimation, one-shot self-supervised learning, Self2Self, Bernoulli sampling, dropout.
\end{IEEEkeywords}

\section{Introduction}
\begin{figure*}[!]
	\begin{centering}
		\includegraphics[width=0.63\textwidth]{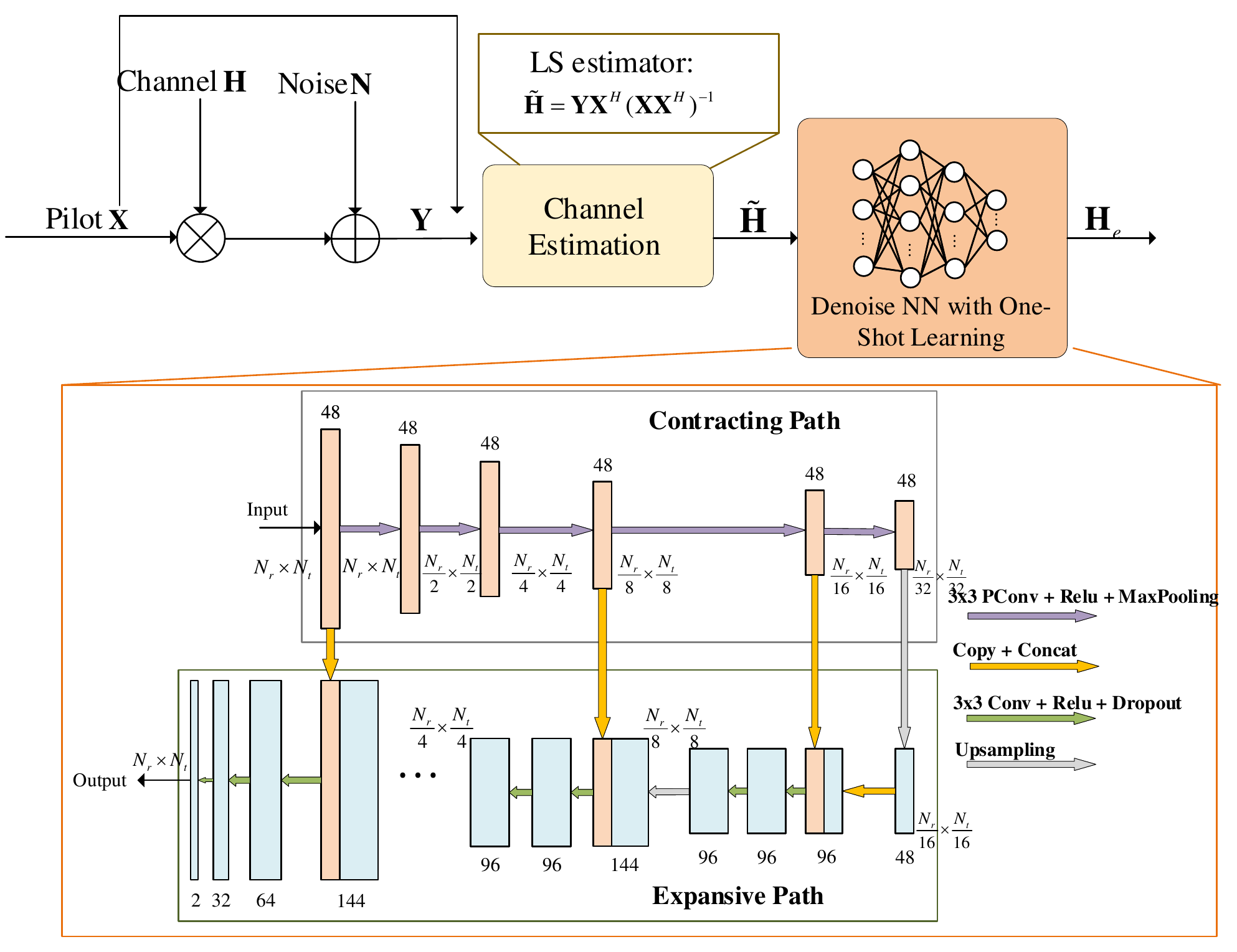}
		\par\end{centering}
	\caption{The structure of the one-shot self-supervised learning framework.}
	\label{framework}
\end{figure*}
Massive multi-input multi-output (MIMO) systems have been widely researched and regarded as a key technique in 5G wireless communications.
To fully exploit the spatial multiplexing gain by multiple antennas, channel state information (CSI) is regularly required for precoding at the base station (BS). Many works consider channel estimation in MIMO systems. Least squares (LS) is a classic estimation approach with low computational complexity but often unsatisfactory performance. Several researchers exploited the sparsity of the channel and designed compressed sensing (CS) based algorithms \cite{Sampling} such as orthogonal matching pursuit (OMP)\cite{OMP}, sparse Bayesian learning (SBL)\cite{SBL} and burst least absolute shrinkage and selection operator (LASSO) \cite{LASSO}. 

CS based algorithms rely on the prior assumption of a sparse structure of MIMO channels. In addition, these techniques often have high computational cost. Recently, deep learning methods have attracted much attention for channel estimation with satisfactory performance under appropriate training and low computational complexity. The authors of \cite{DL1} and  \cite{DL2} designed deep neural network (DNN) based algorithms for channel estimation and a model-driven deep-learning method was proposed in \cite{DL3}. A complex-valued denoising convolution neural network (Cv-DnCNN) was proposed in \cite{DnCNN} to enhance the performance of channel estimation.

The deep learning approaches above are supervised learning methods, which require a large number of samples to optimize the trainable parameters in offline training which render them difficult to implement in practical communication systems. In addition, these networks have a poor generalization ability for example when the SNR is changed. To address the aforementioned problems, some self-supervised learning methods for channel estimation have been proposed, which can be trained online and significantly reduce the number of training samples\cite{DIP1, DIP2}. The authors of \cite{DIP1} proposed a self-supervised learning DNN based on deep image prior (DIP) to denoise the received pilot signals. The main idea of DIP is that the DNN has low impedance to the structured signals and high impedance to the noise, which means that the structured signals are easier learn. Thus, the denoising effect can be achieved by early stopping the training of the DNN. Nevertheless, the performance of DIP is sensitive to the iteration number and it is difficult to determine when to stop the iteration. 

In this paper, we propose a MIMO channel estimation framework based on the one-shot self-supervised learning method Self2Self \cite{Self2Self}, which is not strict on the iteration number. The framework can treat the real-time received pilot signal as the training sample. Thus, it can be directly deployed online without offline training, which can save the training overhead and adapt to variable channels in dynamic environments. The proposed framework consists of the principle estimation module and a denoising module. For the estimation module, we use classic algorithms to obtain a preliminary estimation, which can be viewed as a noisy channel. The denoising module is designed based on a one-shot learning NN with Self2Self, aiming at denoising the noisy channel to obtain a more accurate estimation result. In particular, in the training stage, Bernoulli sampling is performed on the noisy channel to generate training pairs. Blind spot strategy and dropout technique are used to avoid model overfitting to the noisy channel. In the prediction stage, dropout is employed to improve the estimation accuracy.
Simulation results show that our approach achieves better performance than the DIP method and has better generalization ability than the traditional supervised learning method.

\section{Problem Formulation and One-Shot Framework}
\subsection{Problem Formulation}
We consider a MIMO system where the base station (BS) is equipped with $ N_r $ antennas and the receiver is equipped with $ N_t $ antennas in a time-division duplex (TDD) mode. To estimate the downlink channel, we can estimate the uplink channel thanks to channel reciprocity in TDD systems. The user equipment (UE) sends uplink pilots $ \mathbf{X} \in \mathbb{C}^{N_t \times L} $ to the BS, where $ L $ is the length of the pilot. The received signal $ \mathbf{Y} \in \mathbb{C}^{N_r \times L} $ at the BS is given as
\begin{equation}
\mathbf{Y} = \mathbf{H}\mathbf{X} + \mathbf{N},
\end{equation}
where $ \mathbf{H} \in \mathbb{C}^{N_r \times N_t} $ is the channel matrix and $ \mathbf{N} \in \mathbb{C}^{N_r \times L} $ is additive white Gaussian noise (AWGN) with zero mean and variance $ \sigma^2_n $. With the pilots $ \mathbf{X} $ and the received signal $ \mathbf{Y} $, the channel is estimated by an estimator $ \mathcal{F(\cdot)} $ at the BS. The estimated channel is denoted as $ \mathbf{H}_e = \mathcal{F}(\mathbf{X}, \mathbf{Y}) $.

\subsection{Overview of the One-Shot Framework}
In this subsection, we introduce a one-shot self-supervised deep learning framework for channel estimation. Our proposed approach requires no training data, which significantly reduces the training overhead. In addition, the framework can deal with the real-time received pilot signals, which can adapt to the dynamic wireless environments.

The framework is shown in Fig. \ref{framework}, and consists of two modules. The first module is the main channel estimation part using traditional algorithms to obtain a preliminary estimation $ \tilde{\mathbf{H}} $. Specifically, we can use LS estimator, where
\begin{equation}
	\tilde{\mathbf{H}} = \mathbf{Y}\mathbf{X}^H(\mathbf{X}\mathbf{X}^H)^{-1}.
\end{equation}
 $ \tilde{\mathbf{H}} $ can be viewed as a noisy channel and we hope to further improve the estimation performance based on $ \tilde{\mathbf{H}} $, which can be treated as a denoising problem. Thus, we establish the latter module as a channel denoising module, which is designed based on convolutional neural network (CNN) with the one-shot self-supervised learning method Self2Self.
 In particular, the BS received the real-time signal $ \mathbf{Y} $ and pilots $ \mathbf{X} $, and input them to the first module to obtain the preliminary estimation $ \tilde{\mathbf{H}} $. Then we input $ \tilde{\mathbf{H}} $ to the second module, i.e., the denoising NN and train the network online to obtain the final estimation result $ \mathbf{H}_e $.
\section{One-shot Deep Learning Neural Network}
In this section, we first introduce the self-supervised learning method Self2Self for channel denoising and then show the architecture of the denoising NN. Besides, we explain the differences between traditional supervised learning method and our approach.
\subsection{The One-shot Learning Method Self2Self} 
In the one-shot self-supervised learning, we only have one training sample ,i.e., the noisy channel $ \mathbf{\tilde{H}} $ and no corresponding clean channel $ \mathbf{{H}} $.
Thus, the big challenge of the one-shot self-supervised method is over-fitting the noisy channel, which means that the network converges to an identity mapping from  $ \mathbf{\tilde{H}} $ to  $ \mathbf{\tilde{H}} $. Specifically, for channel denoising, we need to minimize the mean square error (MSE) between the estimated channel $ \mathbf{{H}}_e $ and $ \mathbf{H} $. As we know, the denoising NN can be considered as a Bayes estimator and the MSE can be decomposed as the sum of squares of estimation bias and estimation variance, where
\begin{equation}
	\text{MSE} = {bias}^2 + {variance}.
\end{equation} 
When we only have one noisy sample $ \mathbf{\tilde{H}} $ for training, the estimation variance will increase significantly, which causes that the prediction result $ \mathbf{{H}}_e $ converges to the noisy sample $ \mathbf{\tilde{H}} $. Thus, the key point is to avoid the identity mapping and reduce the estimation variance.

To overcome the aforementioned problem, we adopt the self-supervised learning method Self2Self to train the channel denoising NN.
First, we apply the dropout approach in certain layers of NN, which is a common regularization technique in deep learning. It randomly discards some of the neuronal nodes while training, which means that the model structure of the denoising NN in each training epoch is different. In this way, the outputs of these different models have certain degree of statistical independence, which can help to reduce the variance of the prediction result. 

Secondly, to avoid the convergence to an identity mapping, we adopt the strategy of blind spot \cite{Noise2Self}. It refers to that when the network outputs a certain element of channel matrix, the input only contains the information of its surrounding elements instead of the element itself. In general, the channels have strong spatial correlation, which means each element has strong relationships with its surrounding elements. If the noise is spatially independent, the network with blind spot strategy cannot deduce the noise through the surrounding elements but only can predict the channel information related to the surrounding elements. Hence, the noise reduction effect is achieved.
Here we perform Bernoulli sampling on the noisy channel $ \mathbf{\tilde{H}} $ for the blind spot strategy, where
\begin{equation}
	\mathbf{\hat{H}}_{m} = \mathbf{B}_{m} \odot \mathbf{\tilde{H}}, \mathbf{\bar{H}}_{m} = (\mathbf{1}- \mathbf{B}_{m}) \odot \mathbf{\tilde{H}}, \label{Bernoulli}
\end{equation}
where $ \mathbf{B}_{m} \in \mathbb{C}^{N_r \times N_t}$ is the binary Bernoulli matrix with element $ 0 $ or $ 1 $ in the $ m $-th training epoch, $ \odot $ represents multiplication element by element. We denote $ \mathbf{\bar{H}}_{m} $ as the blind channel part which is masked by $ \mathbf{B}_{m} $ and $ \mathbf{\hat{H}}_{m} $ is the unmasked channel part. The input of the NN is $ \mathbf{\hat{H}}_{m} $ and the training loss function is defined as
\begin{equation}
	\sum_{m=1}^{M} \|f_m(\mathbf{\hat{H}}_{m};\boldsymbol{\theta}_m) - \mathbf{\bar{H}}_{m} \|_{\mathbf{B}_m}^2, \label{loss_function}
\end{equation}
where $ M $ is the total number of training epoch, $ f_m(\cdot) $ is the model of the NN in the $ m $-th training epoch, $ \|\mathbf{A}\|_{\mathbf{B}_m} = \|(\mathbf{1}- \mathbf{B}_{m}) \odot \mathbf{A}\|_2^2$. The loss function is measured only on the blind channel part $ \mathbf{\bar{H}}_{m} $ instead of the unmasked input $ \mathbf{\hat{H}}_{m} $, which conforms to the blind spot strategy. 
In addition, when the elements of noise between $ \mathbf{\tilde{H}} $ and $ \mathbf{H} $ are independent and zero mean, the expectation of loss function (\ref{loss_function}) is the same as that of 
\begin{equation}
	\sum_{m=1}^{M} \|f_m(\mathbf{\hat{H}}_{m};\boldsymbol{\theta}_m) - \mathbf{H} \|_{\mathbf{B}_m}^2 + \sum_{m=1}^{M}\|\boldsymbol{\sigma}\|^2_{\mathbf{B}_m}, \label{loss2}
\end{equation}
where $ \boldsymbol{\sigma} $ is the standard deviation of noise between $ \mathbf{\tilde{H}} $ and $ \mathbf{H} $.
The proof is provided in \cite{Self2Self}. Form (\ref{loss2}), we can see that the loss function of training with the paired samples $ \{\hat{\mathbf{H}}, \tilde{\mathbf{H}}\} $ is very related to that of training with the clean labels $ \mathbf{H} $, which infers the rationality of the method.

For the test stage, the dropout technique is also adopted to reduce the prediction variance. We use dropout to generate $ T $ different models $ f_1(\cdot), f_2(\cdot), \ldots, f_T(\cdot) $ to obtain estimated results with certain degree of independence and average them to acquire the recovered channel, where
\begin{equation}
\mathbf{H}_e = \frac{1}{T}\sum_{t=1}^{T}f_t(\mathbf{\hat{H}}_{t};\boldsymbol{\theta}_t).
\end{equation}

\subsection{The Architecture of the One-Shot Denoising NN}
We use the U-Net model \cite{U-Net} for the one-shot learning NN, the architecture of which is shown in Fig. \ref{framework}. The model consists of a contracting path and an expansive path, which are designed for extracting channel features and restoring the original resolution, respectively. Since the input of the NN after Bernoulli sampling can be regarded as degraded channels, the contracting path is set up with partial convolutional (PConv) layers, which are more effectiveness in dealing with degraded signals compared to the traditional convolution layers\cite{PConv}. 
Each PConv layer is followed by a max pooling operation with a stride of $ 2 $ for downsampling.
The number of feature channels expands to 48 from 2 (real and imaginary part of the channel) after the first convolutional layer and then remains the same. The output of the contracting path is the feature with size $ N_r/32 \times N_t/32 \times 48$. 

The expansive path is set up with several blocks. Each block consists of two $ 3 \times 3 $ convolutions and a concatenation operation, and each block is connected with an up-sampling layer with a scaling factor of $ 2 $. The concatenation operation stacks the feature from the up-sampling layer and the corresponding feature in the contracting path to fuse the feature information. The number of feature channels in each output layer of the block is $ 96 $ except for the last block. The size of output feature restores to $ N_r \times N_t \times 2$. In addition, dropout is configured in each convolutional layer of the expansive path. We use rectified linear unit (ReLU) as the activation function in this model.

\subsection{Differences between Supervised Learning and Our Approach}
For channel denoising, in the supervised learning, we first collect a large number of noisy channel samples and corresponding clean channel label as training pairs to optimize the trainable parameters of the NN offline. The optimization target of supervised learning method can be written as 
\begin{equation}
\min\limits_{\boldsymbol{\theta}} \sum_{m}\mathcal{L}(f(\mathbf{\tilde{H}}^{(m)};\boldsymbol{\theta}), \mathbf{H}^{(m)}),
\end{equation}
where $ \mathcal{L}(\cdot) $ is the loss function,  $ \mathbf{\tilde{H}}^{(m)} $ and $ \mathbf{H}^{(m)} $ are the noisy channel samples and clean channel labels in the $ m $-th training epoch, respectively.
After offline training, the NN is deployed to the actual scenario to test online with the parameters of NN fixed. The supervised learning NN works best when the training and test environments are consistent. If the channel statistics in the test environment change, the performance will degrade due to data mismatch.

In the one-shot self-supervised learning, we only have the noisy channel $ \tilde{\mathbf{H}} $ for training. We need to exploit the useful information of $ \mathbf{\tilde{H}} $ and generate training pairs from $ \mathbf{\tilde{H}} $.  The optimization target of our self-supervised learning method can be written as 
\begin{equation}
\min\limits_{\boldsymbol{\theta}} \sum_{m}\mathcal{L}(f(\mathbf{\hat{H}}^{(m)};\boldsymbol{\theta}), \mathbf{\bar{H}}^{(m)}).
\end{equation}
Then the trainable parameters are optimized online with the generated training pairs. In practical applications, the self-supervised one-shot learning NN can be directly deployed online to process the real-time received noisy sample $ \mathbf{\tilde{H}} $. Thus, the proposed self-supervised learning channel estimation approach can adapt to variable channels in dynamic environments.

\section{Simulation Results}
In this section, we verify the performance of the proposed one-shot self-supervised learning framework by simulation results. Normalized mean squared error (NMSE) is used as measurement, where
\begin{equation}
	\text{NMSE} = \dfrac{\|\mathbf{H}_e - \mathbf{H}\|_2^2}{\|\mathbf{H}\|_2^2},
\end{equation} 
where $ \mathbf{H}_e $ is the estimated channel. We generate the channels with '3GPP-3D' model in the QuaDRiGa \cite{Quadriga} simulation platform and set $ N_r  = 64$, $ N_t = 32 $. We use Python as the programming language and use the Pytorch to build the deep learning framework.

In the proposed framework, we use LS estimator for the estimation module. We compare the performance of the proposed Self2Self based self-supervised learning method with that of linear minimum mean square error (LMMSE) estimator, the DnCNN based supervised learning method \cite{DnCNN} and DIP based self-supervised learning method\cite{DIP1}. Besides, we use MMSE estimator as a performance upper bound,  where we assume the channel correlation matrix is known.

\begin{figure}[t]
	\begin{centering}
		\includegraphics[width=0.42\textwidth]{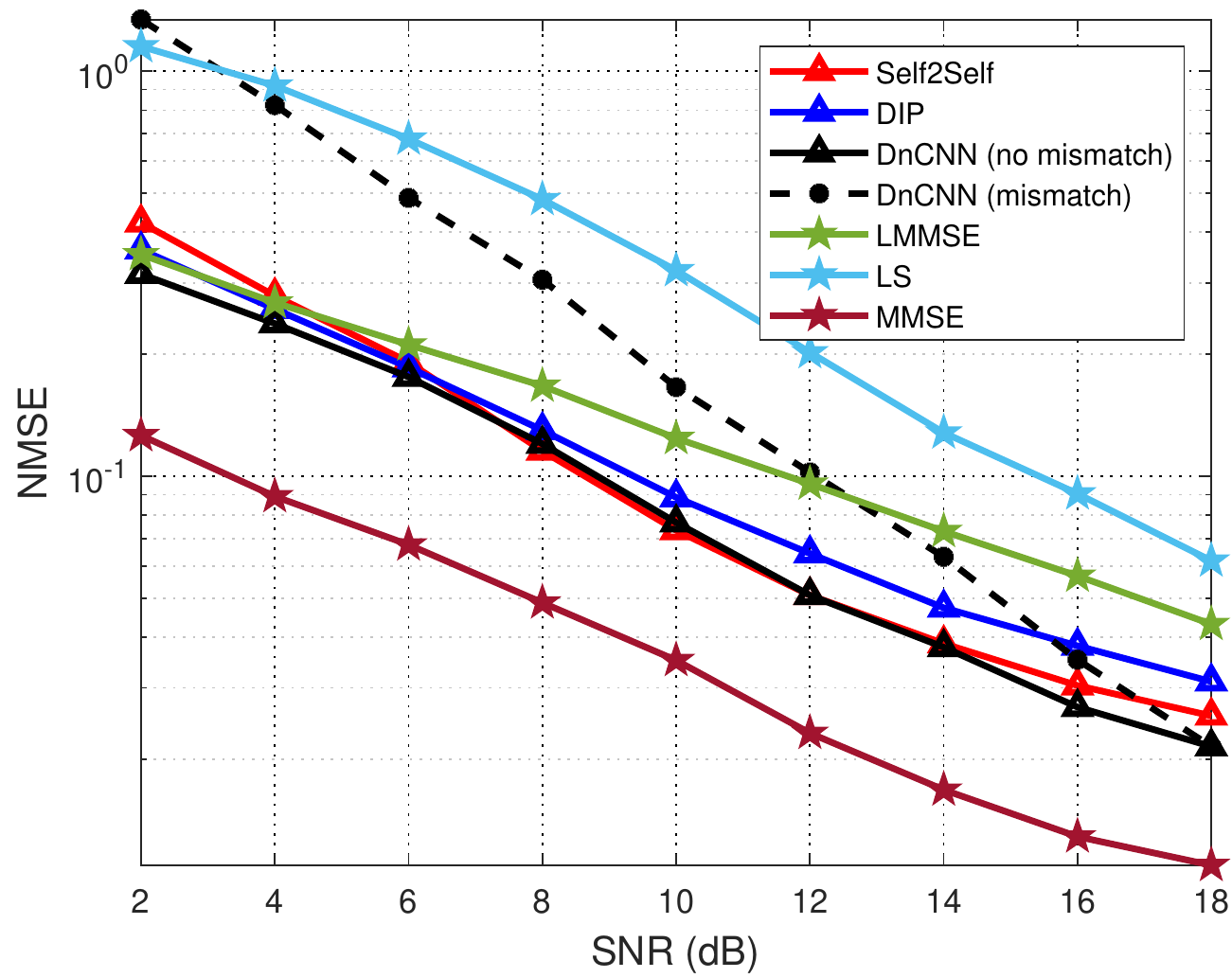}
		\par\end{centering}
	\caption{The NMSE versus the SNR.}
	\label{SNR}
\end{figure}
Fig. \ref{SNR} illustrates the NMSE of the proposed one-shot learning framework and the benchmarks for different values of SNR. For the `DnCNN (no mismatch)', the training and test  SNR configurations are the same; while for the `DnCNN (mismatch)', we train the DnCNN with the configuration $ \text{SNR}$=$ 18 $dB and test it in other SNRs. We can see that the performance of the proposed Self2Self algorithm is much better than the LS estimator, which indicates the effectiveness of the denoising NN. 
Besides, at a high SNR,  the performance of our framework is better than that of the DIP based method and LMMSE estimator. 
Furthermore,  the performance of our proposed self-supervised learning method is very close to that of the `DnCNN (no mismatch)' based supervised method and is better than that of the `DnCNN (mismatch)', which illustrates that our self-supervised learning method has a much better generalization ability than the supervised learning method. The performance gap increases as the SNR decreases, which is because that the mismatch between the training and test stage increases for the supervised learning.

\begin{figure}[t]
	\begin{centering}
		\includegraphics[width=0.42\textwidth]{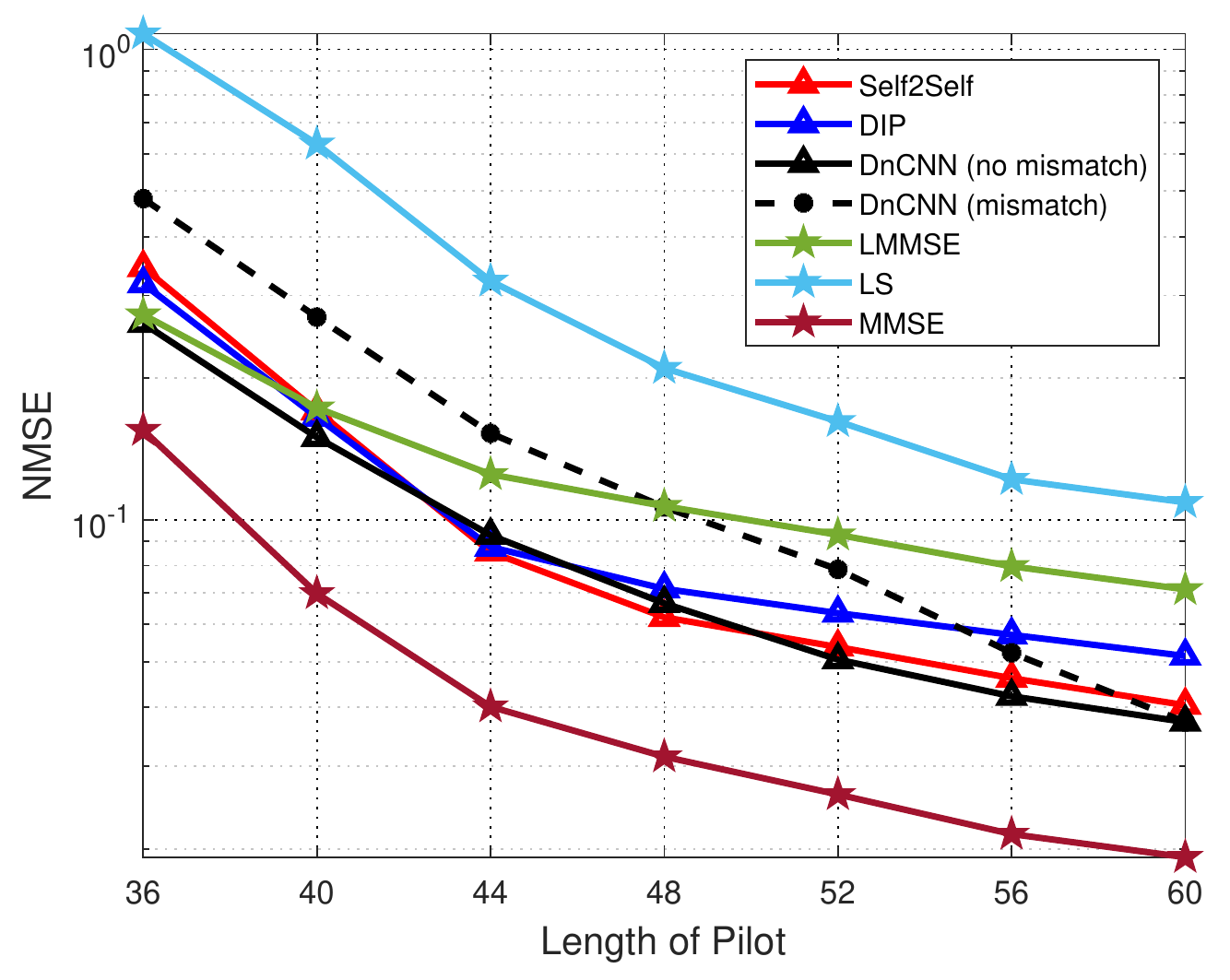}
		\par\end{centering}
	\caption{The NMSE versus the length of pilots.}
	\label{pilot}
\end{figure}
Fig. \ref{pilot} illustrates the NMSE of the proposed one-shot learning framework and the benchmarks for different lengths of pilots. For the `DnCNN (no mismatch)', the pilot length $ L $ in the training and test stage are the same; while for the `DnCNN (mismatch)', we train the DnCNN with $ \text{L}$=$ 60 $ and test it with other pilot lengths. It can be seen that the estimation performance of Self2Self algorithm is much better than that of LS estimator with the same length of pilots. We can see that the NMSE of Self2Self approach with the pilot length of $ 44 $ is less than that of LS estimator with the pilot length of $ 60 $, which demonstrates that our proposed channel estimation method can save the pilots. In addition, our one-shot learning method achieves comparable performance with the `DnCNN (no mismatch)' based supervised learning approach. When there is a mismatch of pilot lengths in train and test stage, our method is much better than the supervised learning, which demonstrates our framework can adapt to the adjusted pilot length.

\begin{figure}[t]
	\begin{centering}
		\includegraphics[width=0.42\textwidth]{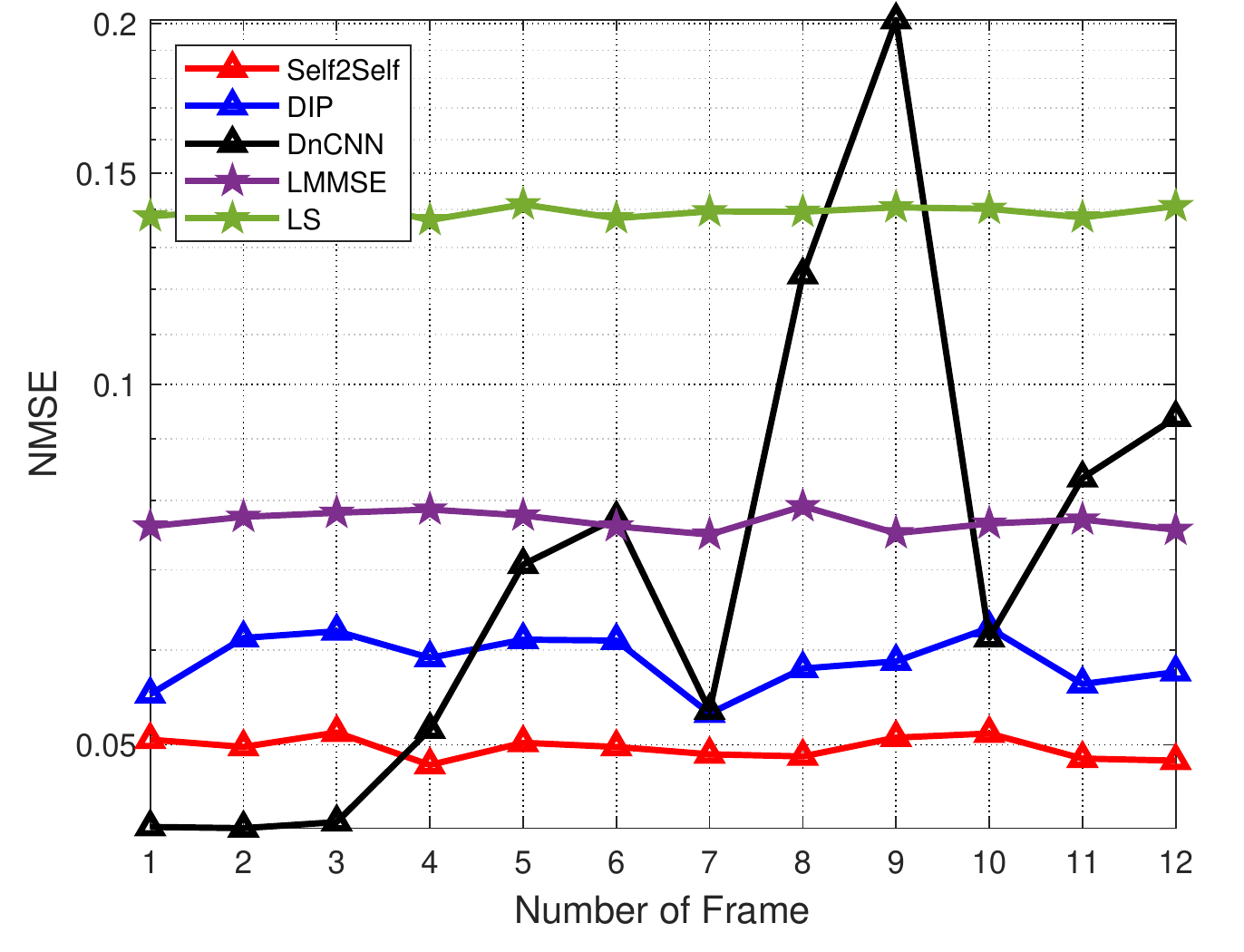}
		\par\end{centering}
	\caption{The NMSE versus the number of frame.}
	\label{Generalize}
\end{figure}
Fig. \ref{Generalize} illustrates the NMSE of the proposed one-shot learning framework and the benchmarks when the channel changes with the frame. We set $ \text{SNR}$=$ 10 $dB and the length of pilot $ \text{L}$=$ 48 $. We simulate the variation of channels by changing the receiver's position. In the first three frames, we keep the receiver's position fixed, where the channel remains unchanged and it is consistent with the training channels for the supervised learning. In this case, the DnCNN based supervised learning method performs well. From the $ 4 $-th frame, we fix the position of BS and move the receiver, which causes that the angle of arrival (AoA) of the LOS path changes. We can see that the performance of the DnCNN based supervised learning method deteriorates due to the mismatch between the training and application environments. At the $ 8 $-th and $ 9 $-th frame, the performance of DnCNN deteriorates dramatically, which is probably because the testing channel varies greatly compared to the training channels. In contrast, the performance of our proposed self-supervised learning method maintains stability, which illustrates that it can adapt to variable channels in dynamic environments.
\section{Conclusion}
In this paper, we proposed a one-shot self-supervised deep learning framework for MIMO channel estimation. It can be directly deployed online without acquiring large numbers of samples for offline training and it is robust to the variable channels in dynamic environments. The framework consists of the traditional estimation module and the denoising module. The denoising NN is designed based on the Self2Self one-shot learning method, where Bernoulli sampling and dropout approaches are employed. The simulation results showed that our proposed self-supervised learning method achieves the comparable performance as the traditional supervised learning method and has a better generalization ability.

\bibliographystyle{IEEEtran}
\bibliography{one-shot}

\end{document}